# The Patchkeeper: An Integrated Wearable Electronic Stethoscope with Multiple Sensors

Hongwei Li[1], Zoran Radivojevic[1], Maja Hedlund[2], Anton Fahlgren[2] and Michael S. Eggleston[2]

[1]Nokia Bell Labs, 21 JJ Thomson Avenue, Cambridge CB3 0FA, United Kingdom
[2]Nokia Bell Labs, 600 Mountain Avenue, Murray Hill, New Jersey, USA
hongwei.3.li@nokia-bell-labs.com

*Abstract*—Many parts of human body generate internal sound during biological processes, which are rich sources of information for understanding health and wellbeing. Despite a long history of development and usage of stethoscopes, there is still a lack of proper tools for recording internal body sound together with complementary sensors for long term monitoring[1]. In this paper, we show our development of a wearable electronic stethoscope, coined "Patchkeeper" (PK), that can be used for internal body sound recording over long periods of time. Patchkeeper also integrates several state-of-the-art biological sensors, including electrocardiogram (ECG), photoplethysmography (PPG), and inertial measurement unit (IMU) sensors. As a wearable device, Patchkeeper can be placed on various parts of the body to collect sound from particular organs, including heart, lung, stomach, and joints etc. We show in this paper that several vital signals can be recorded simultaneously with high quality. As Patchkeeper can be operated directly by the user, e.g. without involving health care professionals, we believe it could be a useful tool for telemedicine and remote diagnostics.

## I. Introduction

As a rich source of physiological information, internal body sounds can provide unique information about processes influencing health including cardiovascular, respiratory, digestive systems, or the health of joints etc.[2], [3], [4], [5], [6]. Sound from the heart and lungs are routinely checked by health care professionals with a stethoscope. Diagnostics usually take place at a stage when symptoms have already developed and seeing a medical professional becomes necessary. Alongside the advancements in various diagnostic techniques, the stethoscope has also progressed to meet the demands of modern medicine and is widely available on the market[7], [8]. These enhancements encompass features like sound visualization, ambient noise reduction, algorithmic diagnostic assistance[9], [10], privacy protection and more[11]. They assist healthcare providers in evaluating the cardiovascular and respiratory systems, as well as in other medical applications like monitoring gastrointestinal tract and bowel sounds[4] or detecting vascular bruits[7]. However, most electronic stethoscopes are designed to be used by health care professionals during in-person patient visits only.

In the meantime, various telemedicine wearables have been developed for daily use to provide health related information remotely, for example heart rate, heart rate variability and blood pressure measurements with smart watch or smart earbuds based on ECG and PPG technologies. However, there is a still a lack of internal body sound capturing devices for more generic health monitoring purposes. Furthermore, some of the early warning signals related to health and wellbeing are based on an episode-type of data occurring under specific conditions and in relatively short periods of time. This puts hard requirements for the device design with long operational time. Therefore we believe it is important to have an easy-to-wear integrated device featuring internal body sounds sensors in addition to complementary opto-electric sensors, capable of autonomous operation in ranges of days. Such a device, aided by modern diagnostic algorithms, would facilitate tracking of heath related information including episode-type of data for cardiovascular, respiratory and digestive systems at home and/or ordinary daily circumstances.

To that end, we developed the Patchkeeper wearable electronic stethoscope, integrated with several state-of-the-art sensing technologies including ECG, PPG and IMU sensors. The overall device's size is 76x52x13 mm with total weight of 56 grams. The device can be supported on most parts of the human body with either detachable adhesive patch electrodes or by use of an elastic strap.

## II. Materials and Methods

The Patchkeeper is designed with the BT840E Bluetooth Low Energy (BLE) module from Fanstel. It contains a nRF52840 chip from Nordic Semiconductor. The nRF52840 SoC is built around the 32-bit ARM® Cortex-M4 CPU with floating point unit running at 64 MHz. In addition to 2.4GHz radio transceiver for BLE, nRF52840 also features multiple digital interfaces that have been used in the PK, including PDM, SPI, I2C and ADC interfaces.

### A. Hardware and mechanical design

We have optimised the design of a single head stethoscope with polymeric diaphragm placed on top of conical bell. The cone shaped head was prototyped by using clay-made mockups and then finally produced by CNC machining in aluminium. The diaphragm and conical bell have been designed to cover a wide band with good sensitivity in the frequency range of 20 Hz to 2 kHz.

Sound received by the stethoscope diaphragm is directly transferred through a hole at the back of the conical bell to a digital MEMS microphone mounted on top of the PCB. The contact is sealed by a soft silicone O-ring as shown in Fig. 1. The second microphone is placed on the same PCB at a different location and is coupled to air only. The second microphone serves for tracking environmental sounds for purpose of noise cancellation and/or privacy protection. Two microphones work as a stereo pair recording continuously into a micro SD card. The microphone used is an

INMP621ACEZ from TDK InvenSense Inc., featuring a PDM digital interface and a dynamic range of 111 dB.

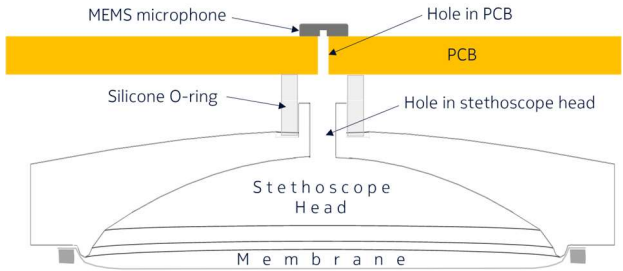

Fig. 1. Schematic of E-stethoscope structures. MEMS microphone is airtight coupled to the back of the stethoscope head via hole in the PCB.

For ECG measurement, we used the ADS1292R chip from Texas Instruments. It is a low-power, dual channel, 24-bit high performance analog front-end (AFE) designed for precise bio-potential measurements. Its dual channel capability enables measurements of ECG and respiration simultaneously. Its exceptional signal quality, compact design and energy efficiency make it ideal for portable, battery-operated devices, enabling continuous, real-time monitoring of vital signs with minimal power consumption.

For PPG measurement, we used a MAXM86161EFD+, a highly integrated optical sensor module from Maxim Integrated, which is designed for advanced health monitoring applications. This compact, ultra-low-power module combines three LEDs, a photo-detector, and a low-noise AFE. It can provide high-resolution measurements of heart rate, blood oxygen saturation, and other vital signs. Its miniature form factor makes it ideal for continuous, real-time health monitoring on wearable devices.

The IMU sensor used is the BMI160 from Bosch Sensortec. This small, low-power, 16-bit chip provides accurate real-time gyroscope and accelerometer data. Combining a 3-axis accelerometer and gyroscope, it's ideal for wearables, offering precise motion tracking with minimal power consumption. Its programmable FIFO buffer ensures efficient integration and responsive motion detection, enhancing various applications.

The PK is equipped with a micro SD card for continuous logging of all sensory data. All data is recorded in binary format with audio and other sensory data in separated files. Each individual sensor's data contains its ID tag and a real-time stamp, and can be parsed with python script later. The PK is powered with a 400mAh lithium polymer battery that is sufficient for 20 hours of continuous data logging. The device can be charged via USB-C port.

Details of the Patchkeeper mechanical design is shown in Fig. 2. ABS material (Acrylonitrile Butadiene Styrene) has been used to fabricate top/bottom covers, mid-frame, and small internal mechanical parts of the PK. In addition, thin and flexible PCBs are used for internal routing and connections of the PPG sensor and ECG electrodes. The mid-frame is used to mechanically secure the stethoscope in the assembly. In addition to the central large hole for stethoscope diaphragm, the bottom cover also has a PPG window and 4 snap button connectors for ECG electrical signal connections with flexible and adhesive patch electrodes. The PK main-body can be mechanically detached (via 4 snap buttons) enabling possible usage of the device without the patch electrodes. A fully assembled PK device with flexible patch electrodes is shown in Fig. 3.

### B. Firmware

The in-house developed Patchkeeper firmware provides comprehensive control over individual components of the system. This includes the ability to turn on and off specific sensors, as well as adjust their sampling mode and frequency. This level of flexibility allows us to customize the PK for various application scenarios, particularly in multimodal sensing applications. In addition to logging all data to the on-board micro SD card, PK is also capable of streaming real time data over BLE for monitoring purpose. Its BLE capability further enables Over-the-Air device firmware update. This feature greatly facilitates firmware updates after deployment, making the device highly suitable for research and exploration projects.

To contribute and support the research and development of home health monitoring technologies, we have made the entire PK design open source. This includes PCB design, mechanical design, firmware, and data parsing script, providing a valuable resource for further advancements in the field. The open source project can be found on GitHub at https://bit.ly/3V1aaS.

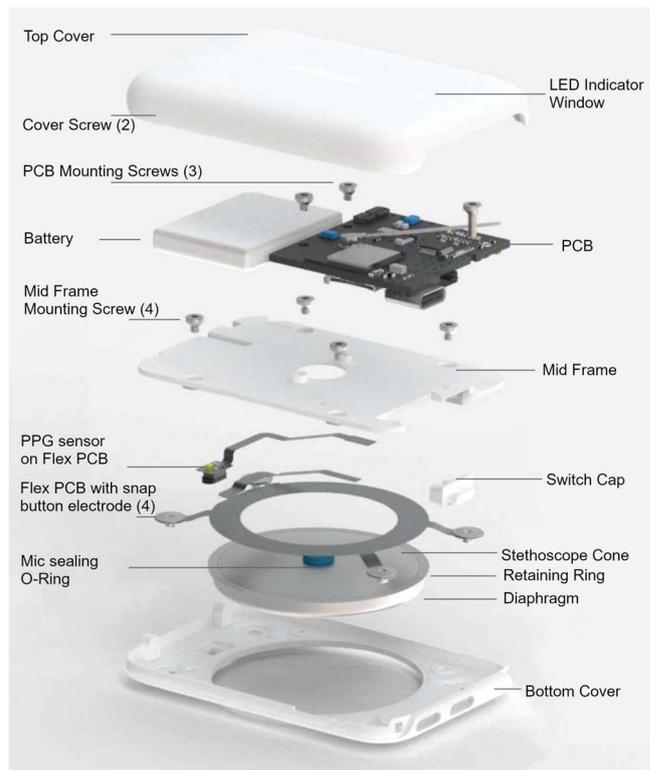

Fig. 2. Details of Patchkeeper mechanical design.

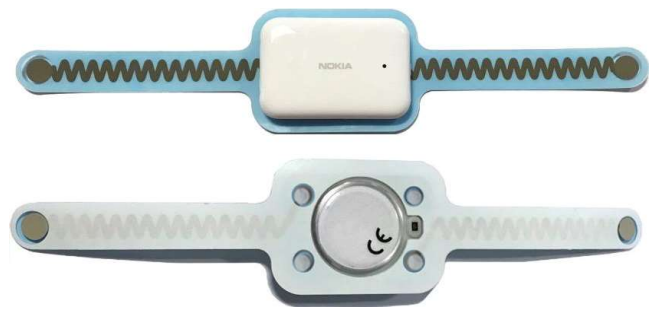

Fig. 3. Picture of full Patchkeeper assembly with flexible and sticky electrodes that can be directly attached to the user skin.

III. RESULTS AND DISCUSSION

The PK is very convenient for tracking human vital signals and internal body sounds including an event-type of data collection (baby's kicking in prenatal phases for example). We conducted simultaneous measurements on a human subject with the all sensors active to ensure reliable functioning of the PK in accordance with the each individual sensor's manufacturer's specifications. The efficacy of the stethoscope was evaluated by recording a phonocardiogram (PCG) e.g. heartbeat sound, in a quiet settings with PK placed around the chest of an adult male. A segment of the recorded sound waveform is depicted in Fig. 4(a) with the ECG signal taken simultaneously in Fig. 4(b). Both the first (S1) and the second (S2) heart sounds can be clearly recognized[12], [13]. By accurately measuring the timing of S1 and S2 and any potential background murmur sound, it is feasible to detect and predict various heart conditions with advanced machine learning algorithms[9], [14], [15]. For the noise cancellation channel, plotted in Fig. 4(a) with offset, it exhibited a very low sound level as expected in a quiet environment. Overall, the sound quality is comparable to our previous prototype, which was successfully utilized for capturing abdominal sounds to extract heart rate[16].

Aiming for multimodal applications, we conducted a test on PK with all sensors activated. During this trial, ECG and respiration (RESP) signals were recorded at a rate of 125 Hz, IMU at 50Hz, and PPG with Green LED channel exclusively at 100 Hz. The sampling rates for both ECG and PPG are suitable for effectively measuring heart rate. These parameters can be adjusted based on particular requirements in real application scenarios.

Fig. 5 shows a segment of 20 seconds of data collected simultaneously from the all sensors recorded on micro SD card. Since all data are gathered using a single device with a synchronized real-time clock, they can be easily and accurately aligned for analysis. Heartbeat signals are clearly visible on both PCG and ECG traces, while RESP signals are clearly visible on the second channel of the ADS1292R and on the IMU gyroscope signals for y and z axes. The respiration induced movement is also visible in IMU accelerometer data (marked as az).

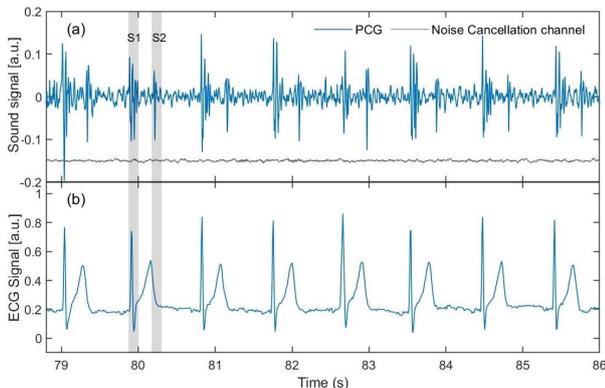

Fig. 4. Simultaneous recording of heart activity by E-stethoscope (a) and ECG (b). Positions of S1 and S2 signals from heart sound trace are roughly marked with gray bars.

For the PPG sensor, although the signal from the green LED channel appears noisy, a detailed analysis reveals that it contains components from both heartbeat and respiration-induced movement. With the use of a proper algorithm, heart rate can still be obtained[17].

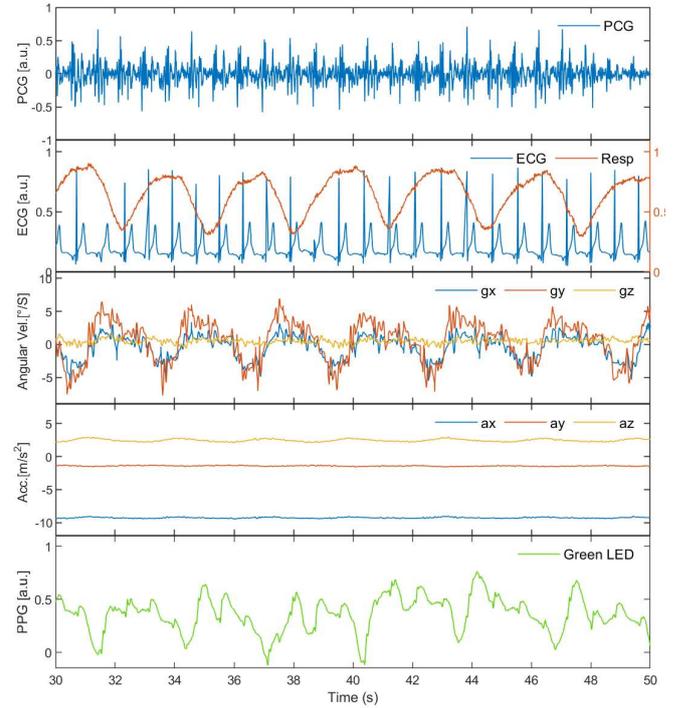

Fig. 5. Example of 20 seconds long data from all sensors simultaneously recorded on micro SD card.

IV. CONCLUSION

We presented our recent development of Patchkeeper, a wearable electronic stethoscope integrated with multiple sensors suitable for multimodal long-term monitoring. As a comfortable and wearable device, PK can be conveniently utilized outside of clinical environments, allowing for the recording of internal body sounds over extended periods to capture episode-type of data. Equipped with sensors such as ECG, PPG, and IMU, a single Patchkeepr can serve as a platform with multimodal sensors suitable for tracking outpatients released from the hospital or users at home environments. It's easy to use and provides rich dataset enabling meaningful applications for both users and health care professionals.

Moreover, the PK is well suited for various research endeavors. We have used the PK for tracking abdominal body sounds in subjects exposed to different levels of daily stresses[18] and for heart rate information extraction[16]. Furthermore, we have also employed PK to quantify the personality traits of dogs[19]. Additionally, the PK can be utilized for applications such as monitoring premature babies or for smart home applications including intelligent home monitoring and predictive maintenance.

## REFERENCES

[1] J.-Y. Yoo et al., 'Wireless broadband acousto-mechanical sensing system for continuous physiological monitoring', Nature Medicine, vol. 29, no. 12, pp. 3137–3148, Dec. 2023, doi: 10.1038/s41591-023-02637-5.

[2] J. J. Seah, J. Zhao, D. Y. Wang, and H. P. Lee, 'Review on the Advancements of Stethoscope Types in Chest Auscultation', Diagnostics, vol. 13, no. 9, 2023, doi: 10.3390/diagnostics13091545.

[3] C. Pinto, D. Pereira, J. Ferreira-Coimbra, J. Português, V. Gama, and M. Coimbra, 'A comparative study of electronic stethoscopes for cardiac auscultation', in 2017 39th Annual International Conference of the IEEE Engineering in Medicine and Biology Society (EMBC), 2017, pp. 2610–2613. doi: 10.1109/EMBC.2017.8037392.

[4] H. Ashrafian et al., 'Metabolomics: The Stethoscope for the Twenty-First Century.', Med Princ Pract, vol. 30, no. 4, pp. 301–310, 2021, doi: 10.1159/000513545.

[5] R. S. Vasudevan et al., 'Persistent Value of the Stethoscope in the Age of COVID-19.', Am J Med, vol. 133, no. 10, pp. 1143–1150, Oct. 2020, doi: 10.1016/j.amjmed.2020.05.018.

[6] J. Lemejda, M. Kajor, D. Grochala, M. Iwaniec, and J. E. Loster, 'Synchronous Auscultation of Temporomandibular Joints Using Electronic Stethoscopes', in 2020 IEEE XVIth International Conference on the Perspective Technologies and Methods in MEMS Design (MEMSTECH), 2020, pp. 146–149. doi: 10.1109/MEMSTECH49584.2020.9109447.

[7] P.-W. L. Frank and M. Q.-H. Meng, 'A low cost Bluetooth powered wearable digital stethoscope for cardiac murmur', in 2016 IEEE International Conference on Information and Automation (ICIA), 2016, pp. 1179–1182. doi: 10.1109/ICInfA.2016.7831998.

[8] N. Tharapecharat et al., 'Digital stethoscope with processing and recording based on cloud', in 2023 15th Biomedical Engineering International Conference (BMEiCON), 2023, pp. 1–4. doi: 10.1109/BMEiCON60347.2023.10322053.

[9] S. B. Shuvo, S. S. Alam, S. U. Ayman, A. Chakma, P. D. Barua, and U. R. Acharya, 'NRC-Net: Automated noise robust cardio net for detecting valvular cardiac diseases using optimum transformation method with heart sound signals', Biomedical Signal Processing and Control, vol. 86, p. 105272, 2023, doi: https://doi.org/10.1016/j.bspc.2023.105272.

[10] M. Z. Belmecheri, M. Ahfir, and I. Kale, 'Automatic heart sounds segmentation based on the correlation coefficients matrix for similar cardiac cycles identification', Biomedical Signal Processing and Control, vol. 43, pp. 300–310, 2018, doi: https://doi.org/10.1016/j.bspc.2018.03.009.

[11] M. Abdallah, R. Hassan, and E. Jeremy, 'A novel smart stethoscope for health-care providers', in 2017 IEEE MIT Undergraduate Research Technology Conference (URTC), 2017, pp. 1–4. doi: 10.1109/URTC.2017.8284208.

[12] R. Prasad, G. Yilmaz, O. Chetelat, and M. Magimai.-Doss, 'Detection Of S1 And S2 Locations In Phonocardiogram Signals Using Zero Frequency Filter', in ICASSP 2020 - 2020 IEEE International Conference on Acoustics, Speech and Signal Processing (ICASSP), 2020, pp. 1254–1258. doi: 10.1109/ICASSP40776.2020.9053155.

[13] S. Banerjee, M. Mishra, and A. Mukherjee, 'Segmentation and detection of first and second heart sounds (Si and S2) using variational mode decomposition', in 2016 IEEE EMBS Conference on Biomedical Engineering and Sciences (IECBES), 2016, pp. 565–570. doi: 10.1109/IECBES.2016.7843513.

[14] B. Omarov, A. Batyrbekov, A. Suliman, B. Omarov, Y. Sabdenbekov, and S. Aknazarov, 'Electronic stethoscope for detecting heart abnormalities in athletes', in 2020 21st International Arab Conference on Information Technology (ACIT), 2020, pp. 1–5. doi: 10.1109/ACIT50332.2020.9300109.

[15] M. M. S. J. P. P. Careena and P. Arun, 'Statistically significant feature-based heart murmur detection and classification using spectrogram image comparison of phonocardiogram records with machine learning techniques', Australian Journal of Electrical and Electronics Engineering, vol. 0, no. 0, pp. 1–15, 2024, doi: 10.1080/1448837X.2024.2312491.

[16] J. Stuchbury-Wass, E. Bondareva, K. -J. Butkow, S. Šćepanović, Z. Radivojevic, and C. Mascolo, 'Heart Rate Extraction from Abdominal Audio Signals', in ICASSP 2023 - 2023 IEEE International Conference on Acoustics, Speech and Signal Processing (ICASSP), Jun. 2023, pp. 1–5. doi: 10.1109/ICASSP49357.2023.10096600.

[17] A. Ferlini, A. Montanari, C. Min, H. Li, U. Sassi, and F. Kawsar, 'In-Ear PPG for Vital Signs', IEEE Pervasive Computing, vol. 21, no. 1, pp. 65–74, 2022, doi: 10.1109/MPRV.2021.3121171.

[18] E. Bondareva et al., 'Stress Inference from Abdominal Sounds using Machine Learning', in 2022 44th Annual International Conference of the IEEE Engineering in Medicine & Biology Society (EMBC), Jul. 2022, pp. 1985–1988. doi: 10.1109/EMBC48229.2022.9871165.

[19] L. Meegahapola, M. Constantinides, Z. Radivojevic, H. Li, D. Quercia, and M. S. Eggleston, 'Quantified Canine: Inferring Dog Personality From Wearables', in Proceedings of the 2023 CHI Conference on Human Factors in Computing Systems, in CHI '23. New York, NY, USA: Association for Computing Machinery, 2023. doi: 10.1145/3544548.3581088.